\documentclass{PoS}
\usepackage{lineno}  % for line numbering during review
\title{$CP$ violation in $B_s^0$ mixing at LHCb}

\ShortTitle{$CP$ violation in $B_s^0$ mixing at LHCb}

\author{\speaker{Yasmine Amhis}\thanks{The author would like to thank the LHCb Collaboration, the HQL 2012 organisers and the EPFL group.}\\
        On behalf of the LHCb Collaboration\\
        \'Ecole Polytechnique F\'ed\'erale de Lausanne\\
        E-mail: \email{Yasmine.Amhis@epfl.ch}}

\abstract{The {\it CP} violating phase $\phi_s$ can be measured in the interference between mixing and decay of $B_s^0$ mesons decaying to $CP$ eigenstates. The phase $\phi_s$ the decay width difference $\Delta\Gamma_s$ and the average decay width $\Gamma_s$ have been measured at LHCb, using the full 1 fb$^{-1}$ of $pp$ collisions data at a centre-of-mass energy $\sqrt{s} $ = 7 TeV collected during the 2011 LHC run. }
\FullConference{The XIth International Conference on Heavy Quarks and Leptons,\\
		June 11-15, 2012\\
		Prague, Czech Republic}

\begin{document}
\section{Introduction}
These last years, the decay $B_s^0\rightarrow J/\psi\phi$ made a name for itself. It is nowadays,  known as the golden mode to measure $CP$ violation in the $B_s^0$ system. Lately the decay $B_s^0\rightarrow J/\psi\pi\pi$ has also gained some popularity. In the Standard Model the $CP$ violating phase if sub-leading penguin contributions are neglected is predicted to be $\phi_s\simeq -2\beta_s$, 
where $\beta_s = arg(-V_{ts}V_{tb}^{*}/ V_{cs}V_{cb}^{*})$~\cite{Dighe, Dunietz}. The indirect determination via global fits to experimental data gives $2\beta_s = 0.036^{+0.0016}_{-0.0015}$ rad~\cite{Charles2011, Lenz2006,Lenz2011}, Contributions from physics beyond the Standard Model may affect the measured value of $\phi_s$ \cite{Ligeti2006,Ball2006, Lenz2009,Fleisher2006 ,Nierste2008}.
Already, during the summer 2011 LHC run, LHCb collected about $0.4$ fb$^{-1}$ of $pp$ collisions at $\sqrt{s}=$ 7 TeV. Using both decay channels $B_s^0\rightarrow J/\psi\phi$ and $B_s^0\rightarrow J/\psi\pi\pi$, LHCb measured the most precise value of $\phi_s$~\cite{PhisLHCbPRL2011,PhisLHCbPiPi2011}. These measurements were updated using 1 fb$^{-1}$ of data, where 21200 and 7400 $B_s^0\rightarrow J/\psi\phi$ and $B_s^0\rightarrow J/\psi\pi\pi$  candidates where selected. These results are presented in these proceedings. 
A detailed description of these analyses can be found in dedicated published papers and conference reports~\cite{PhisLHCbCONF,  LHCbSignDeltaGammas, PhisLHCbPiPi2012}.
%%-----------------------------------------------------------------------------
%%-----------------------------------------------------------------------------

\section{$B_s^0\rightarrow J/\psi\phi$ analysis}
\begin{figure}[h]
\centering
\includegraphics[scale=0.35]{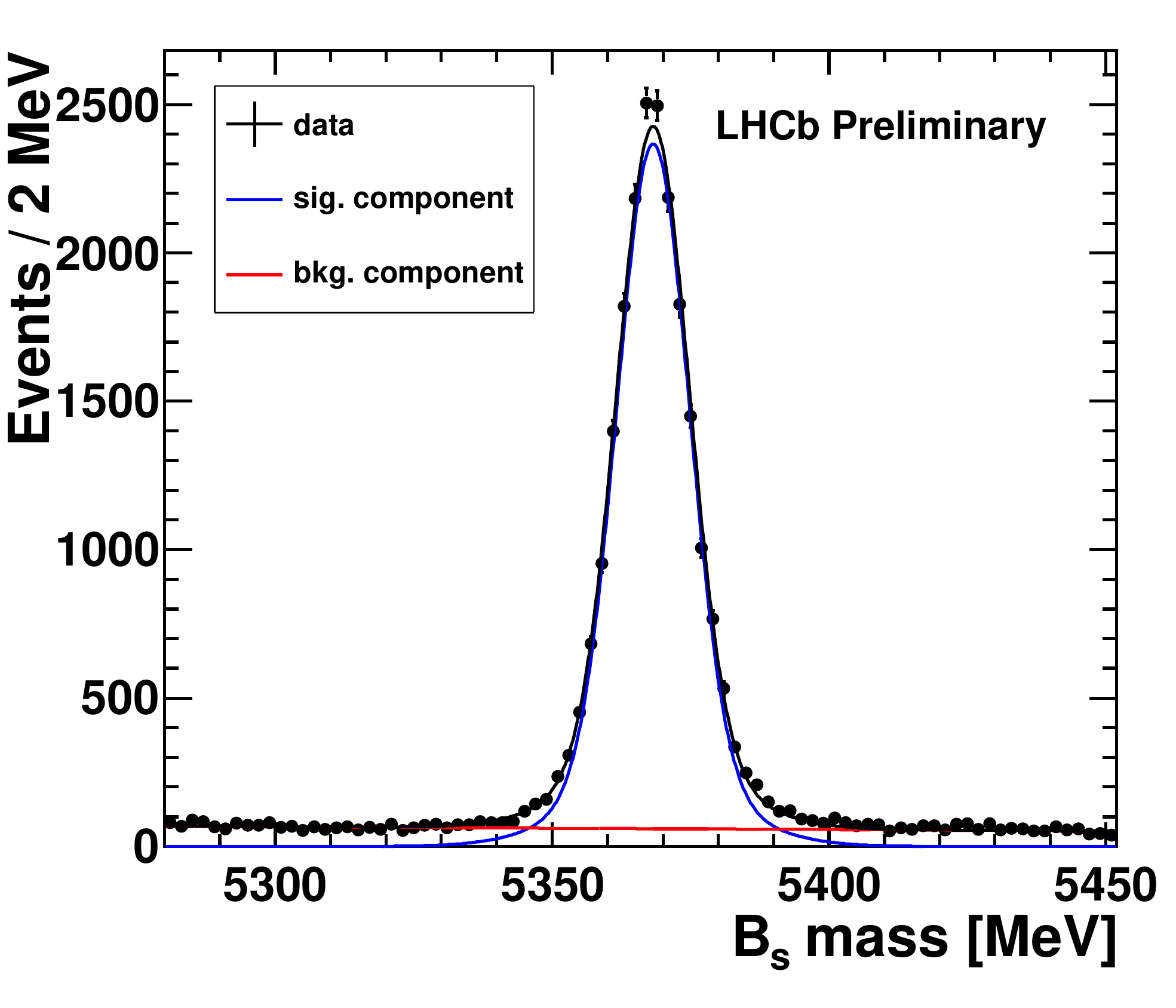}
\caption{Invariant mass distribution of selected $B_s^0\rightarrow J/\psi\phi$ candidates. The background
is shown as the horizontal (red dotted) line.}
\label{fig:Bs2JpsiphiMass}
\end{figure}

The updated  $B_s^0\rightarrow J/\psi\phi$ analysis uses the same event selection as described in Ref.~\cite{PhisLHCbPRL2011}. However, the trigger conditions in 2011 where such that  a decay time biasing cut was introduced in the second half of the data taking. Therefore a dedicated acceptance
is used to correct for this effect. To improve the description of the data a per-event estimation of the decay time resolution is included in the analysis.  To maximise the sample purity, prompt background events are removed by requiring that each $B_s^0$ candidate has a decay time higher than 0.3 ps. The final selected  sample contains about 21200 $B_s^0\rightarrow J/\psi\phi$ candidates as shown in Fig.~\ref{fig:Bs2JpsiphiMass}. 
The strategy for the optimisation and calibration of the flavour tagging is described in detail in Ref.~\cite{OSTagging}. The "opposite-side" (OS) 
flavour tagger exploits the decay of the other $b$-hadron produced in the event and uses four different signatures,  namely high $p_T$ muons, electrons and kaons, and the charge of an inclusive reconstructed secondary vertex. The combination of these taggers provides an estimated per-event mistag probability. 
The OS is calibrated using the $B^+\rightarrow J/\psi K^+$ decays as they do not oscillate.  The effective average mistage probability is  $\omega=(36.81\pm 0.18\pm0.74)\%$. The signal tagging efficieny is $\epsilon_{\textrm{tag}} = (32.99\pm 0.33)\%$. Thus the effective tagging efficiency is $\epsilon_{\textrm{tag}} {\cal D}^2= (2.29 \pm 0.07\pm 0.26) \%$, where ${\cal D}$ is the dilution, definded as ${\cal D} = (1-2\omega)$. The effect of a possible small difference in mistag
probability between both flavours of the $B_s^0$ were estimated to be negligible.  The uncertainties from flavour tagging calibration are included 
in the statistical uncertainties of the physics parameters presented in the next section by allowing the tagging calibration parameters to vary in the final fit within their uncertainties.  
To account for the finite decay time resolution of detector, the Probability Density Functions (PDFs) used in the fit are convolved with a Gaussian function. 
The witdh of the Gaussian is $S_{\sigma_{t}}.\sigma_t$, where the $\sigma_t$ is the estimated per-event decay time resolution.  The scale factor $S_{\sigma_{t}}$ is allowed to vary within its uncertainties in the fit. The effective average decay time resolution is approximatively 45 fs. 
The triggers used in this analysis exploits the signature of $J/\psi\rightarrow \mu\mu$ including decay time biasing cuts. 
The effect of the trigger selection is measured using a set of similar presecaled trigger lines, that do not require the decay time biasing cut. 
A non-parametric description of the acceptance is used in the likelihood fit.  Using simulated events, an acceptance at high lifetimes attributed to the reduction of the track finding efficiency for tracks originating from displaces vertices produced far from the beam line was observed. 
A correction is determined using simulation  and found to be 0.0112 $\pm$ 0.0013 ps$^{-1}$ on $\Gamma_s$. This correction is also accounted for in the final fit. 
The decay angle acceptance is obtained using simulated events and taken into account in the fit. Differences between simulated and observed kaon momentum spectra as well as the limited size of the sample are used to derive corresponding systematic uncertainties. 
%%-----------------------------------------------------------------------------

\begin{figure}[h]
\centering
\includegraphics[scale=0.40]{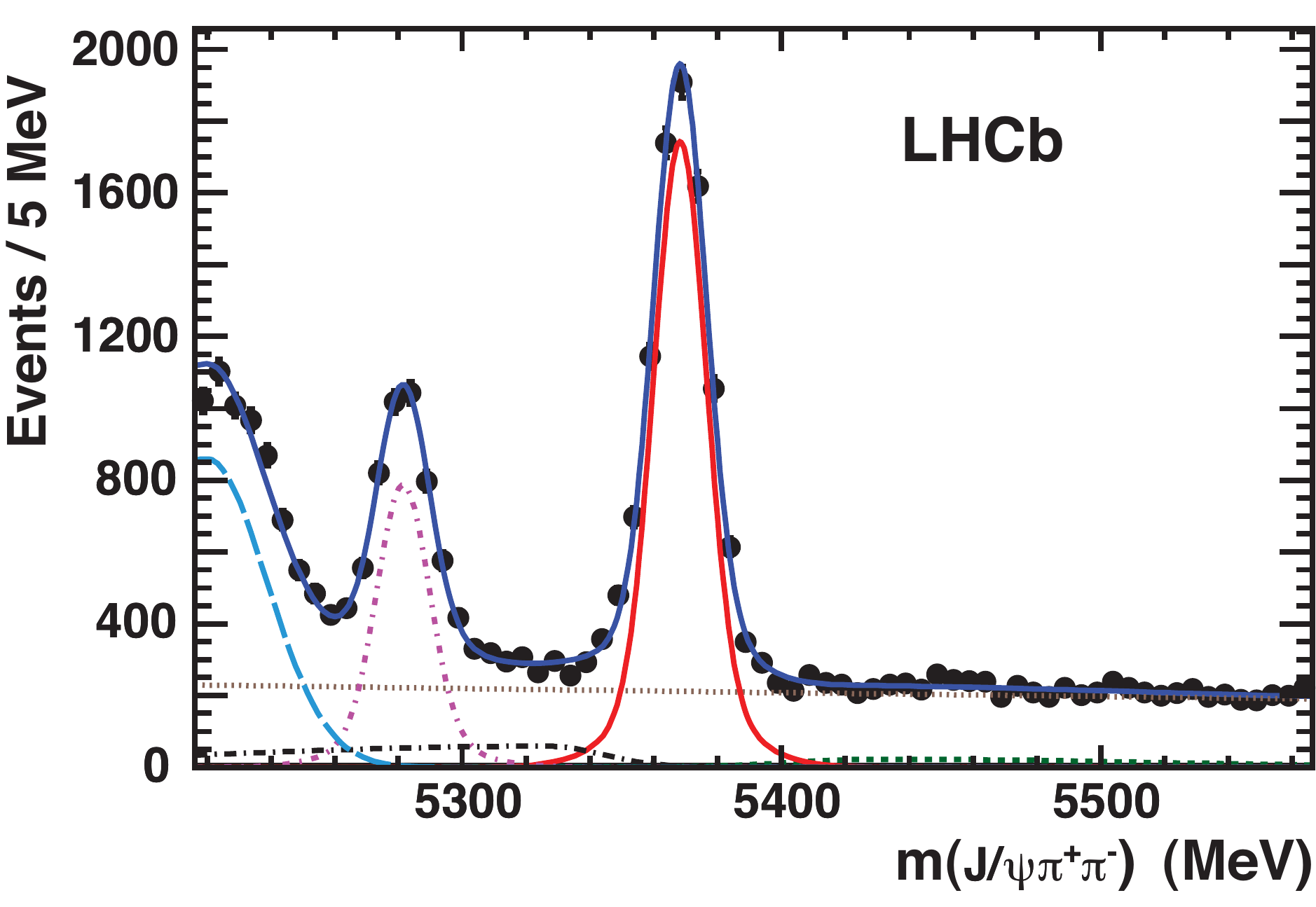}
\caption{Invariant mass distribution of selected
$B_s^0\rightarrow J/\psi\pi\pi$ candidates. The signal is shown as the red solid line. Backgrounds are
combintorial (brown dotted) and $B_s^0\rightarrow J/\psi\pi\pi$ (black long-dot). Other backgrounds are defined
in~\cite{PhisLHCbPiPi2012} but are irrelevant as the analysis only uses the data above a mass of 5346~MeV/$c^2$.}
\label{fig:Bs2JpsipipiMass}
\end{figure}
%%-----------------------------------------------------------------------------
\section{$B_s^0\rightarrow J/\psi\pi\pi$ analysis}
The measurement of $\phi_s$ in $B_s^0\rightarrow J/\psi\pi\pi$ using 1 fb$^{-1}, $ is now published in Ref.~\cite{PhisLHCbPiPi2012}.
In terms of event selection, trigger requirements and tagging information, the analysis strategy is very close to the previous published result~\cite{PhisLHCbPiPi2011}. The main difference is that for the update the $\pi\pi$ spectrum was extended to [775-1550] MeV$/c^{2}$.  A dedicated modified Dalitz analysis was performed to study the resonance and non resonant contributions to  the $\pi\pi$ system. It was shown that this spectrum is dominated by a $CP$-odd component via  the $f_0(980)$ meson decay.  About 7400 signal events are selected and shown in Fig.\ref{fig:Bs2JpsipipiMass}. 
\begin{table}[t]
\begin{center}
\begin{tabular} {|c|c|c|c|}
\hline
Parameter                  & Value & Stat. & Syst. \\ 
\hline\hline
$\Gamma_s$ [ps$^{-1}$]        & 0.6580 & 0.0054 & 0.0066\\
$\Delta \Gamma_s$ [ps$^{-1}$] & 0.116 & 0.018 & 0.006\\
$|A_{\perp}(0)|^2$              & 0.246 & 0.010 & 0.013\\
$|A_{0}(0)|^2$                  & 0.523 & 0.007 & 0.024\\
$F_\mathrm{S}$                    & 0.022 & 0.012 & 0.007\\
$\delta_{\perp}$  [rad]         & 2.90  & 0.36  & 0.07\\
$\delta_{\parallel}$  [rad]         & \multicolumn{2}{c|}{[2.81, 3.47]}  & 0.13\\
$\delta_s$     [rad]            & 2.90  & 0.36  &  0.08\\
$\phi_s$         [rad]          & -0.001  & 0.101  &  0.027\\ \hline
\end{tabular}
\end{center}
\caption{Results for the physics parameters and their
statistical and systematic uncertainties. We quote a 68\% C.L. interval for $\delta_{\parallel}$, as described in the text. \label{tab:final+unc.}}
\end{table}

%%-----------------------------------------------------------------------------
\section{Results}
\begin{figure}[h]
\centering
\includegraphics[scale=0.35]{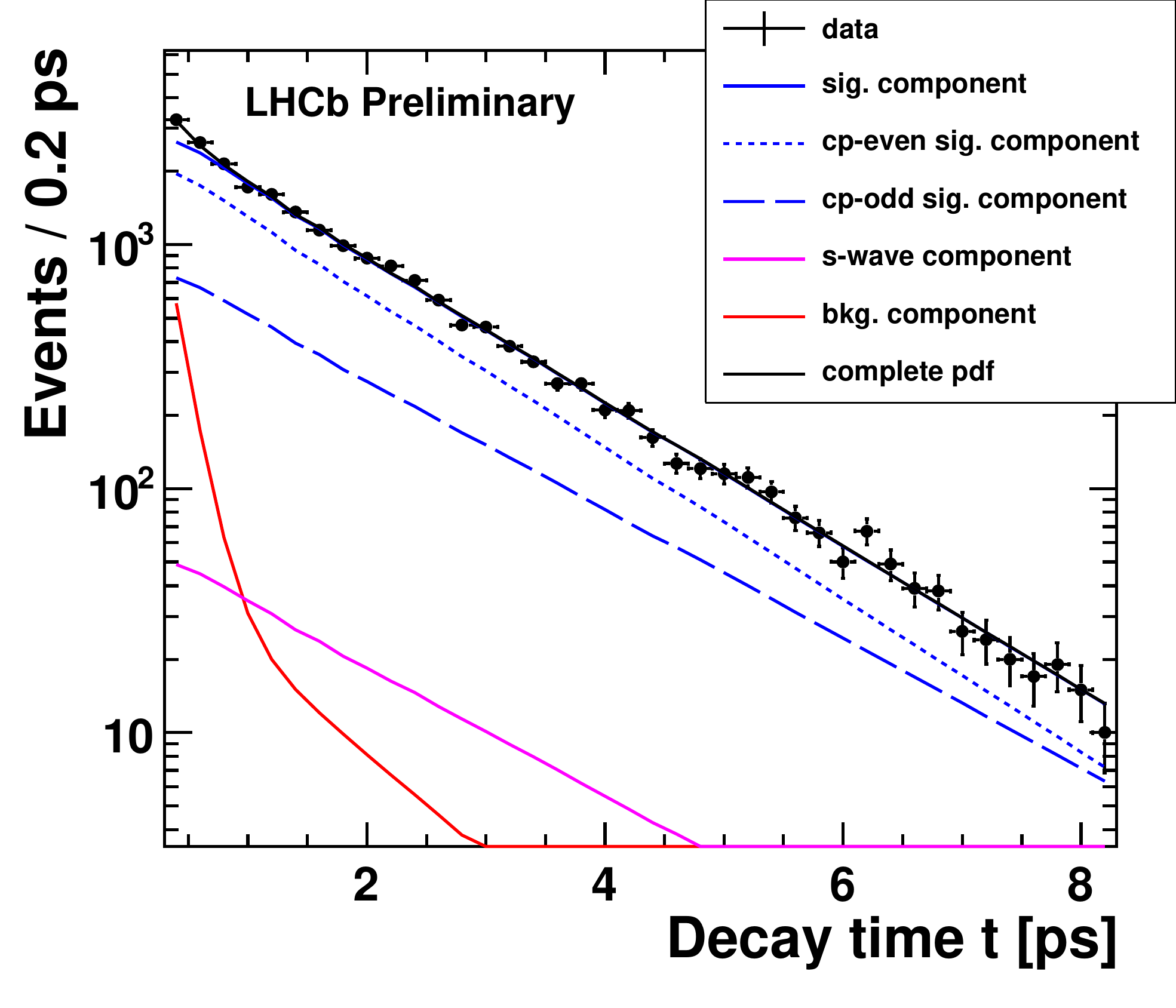}
\includegraphics[scale=0.35]{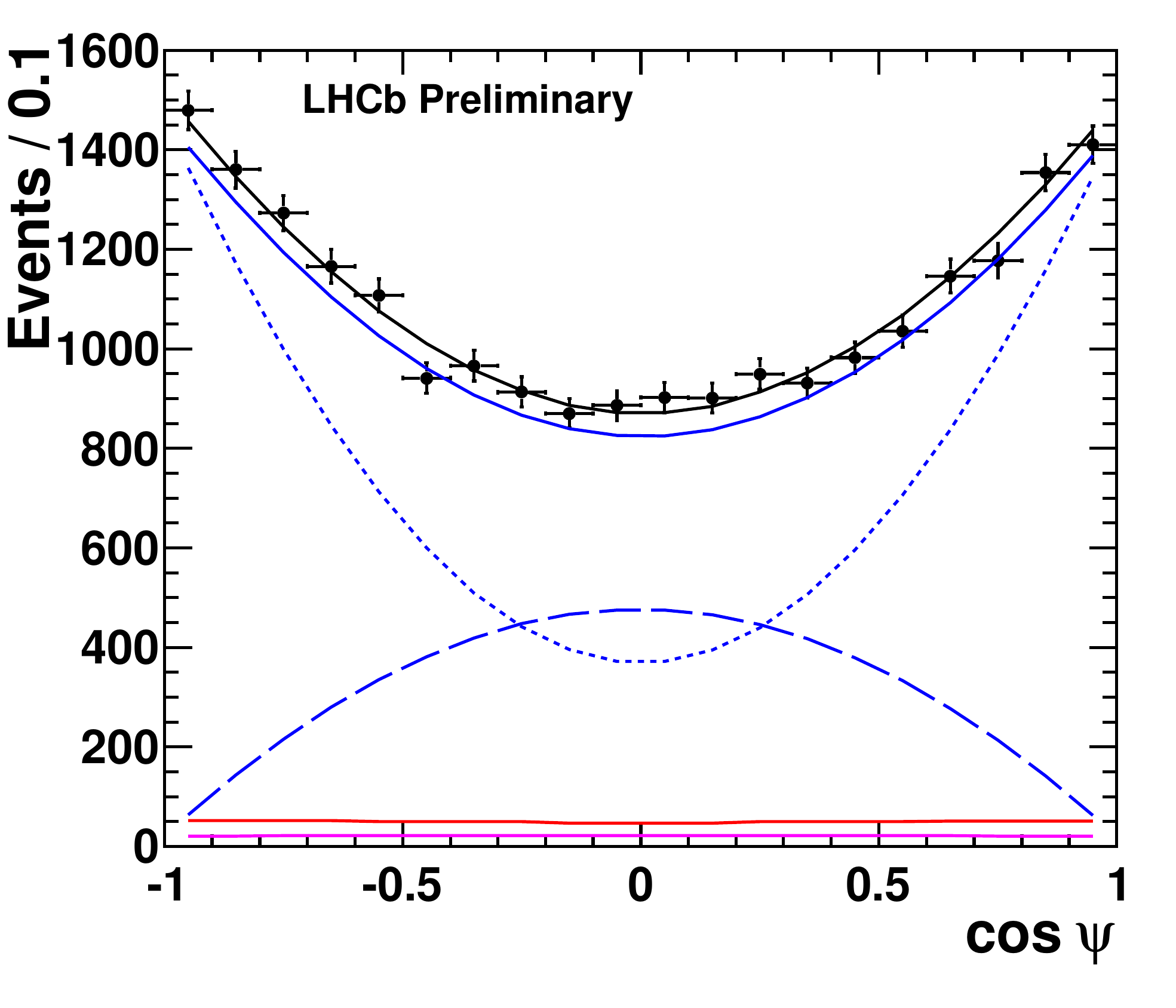}
\includegraphics[scale=0.35]{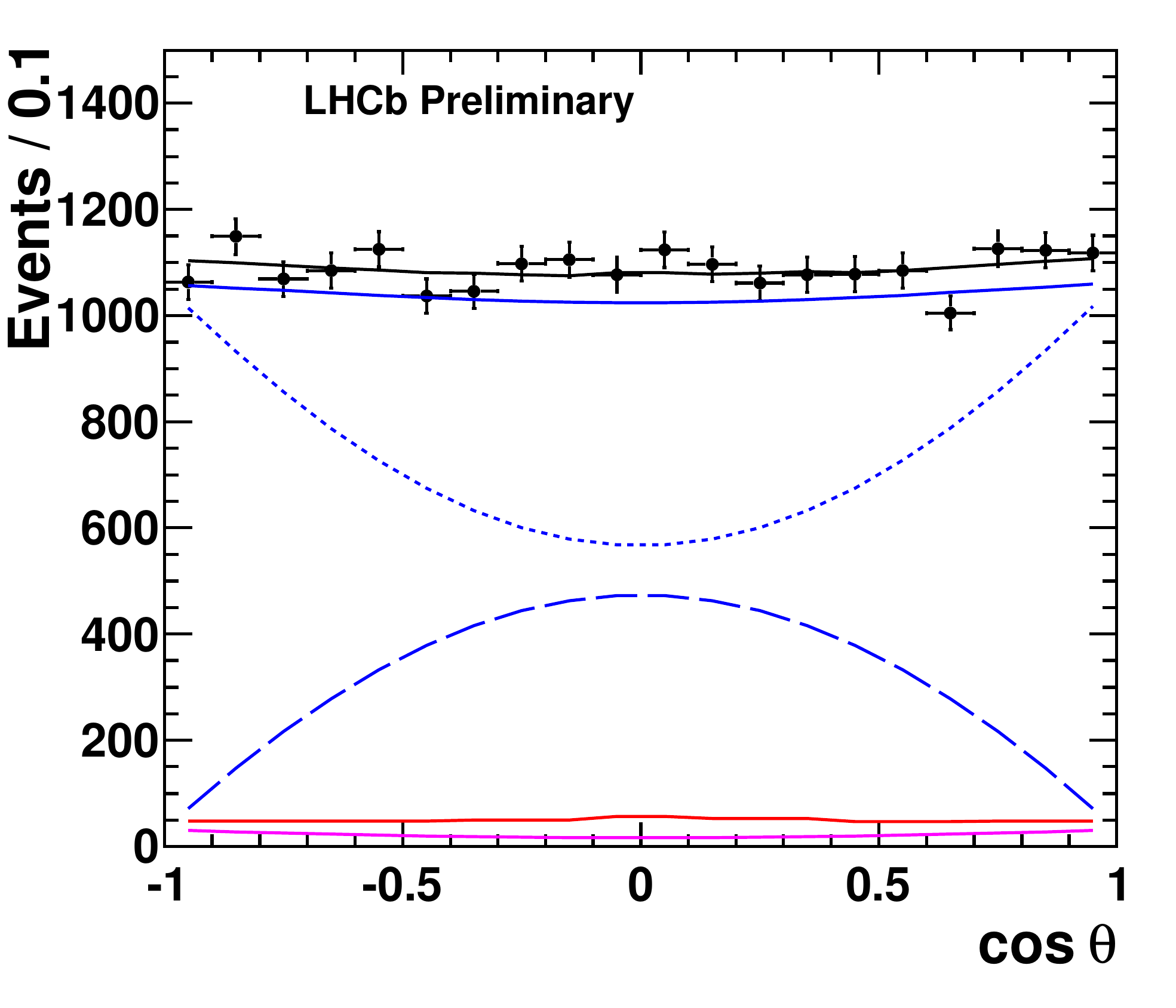}
\includegraphics[scale=0.35]{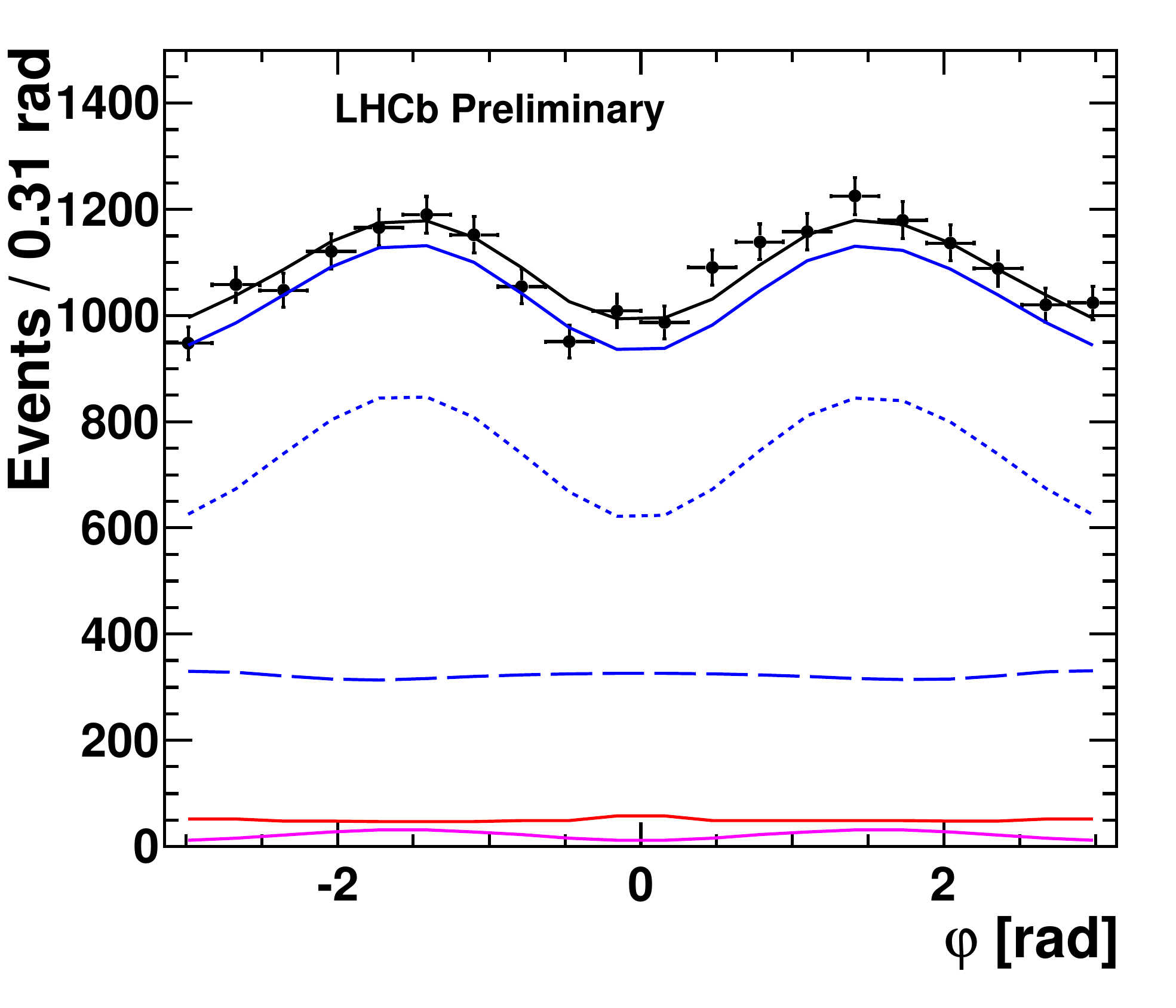}
\caption{Data points overlaid with fit projections for the decay time and transversity angle
distributions in a mass range of $\pm$20 MeV/$c^2$ around the reconstructed $B_s^0$ mass. The total
fit result is represented by the black line. The signal component is represented by the solid
blue line; the dashed and dotted blue lines show the $CP$-odd and $CP$-even signal components
respectively. The S-wave component is represented by the solid pink line. The background
component is given by the red line.}
\label{fig:projections}
\end{figure}

\begin{figure}[h]
\center
\includegraphics[scale=0.41]{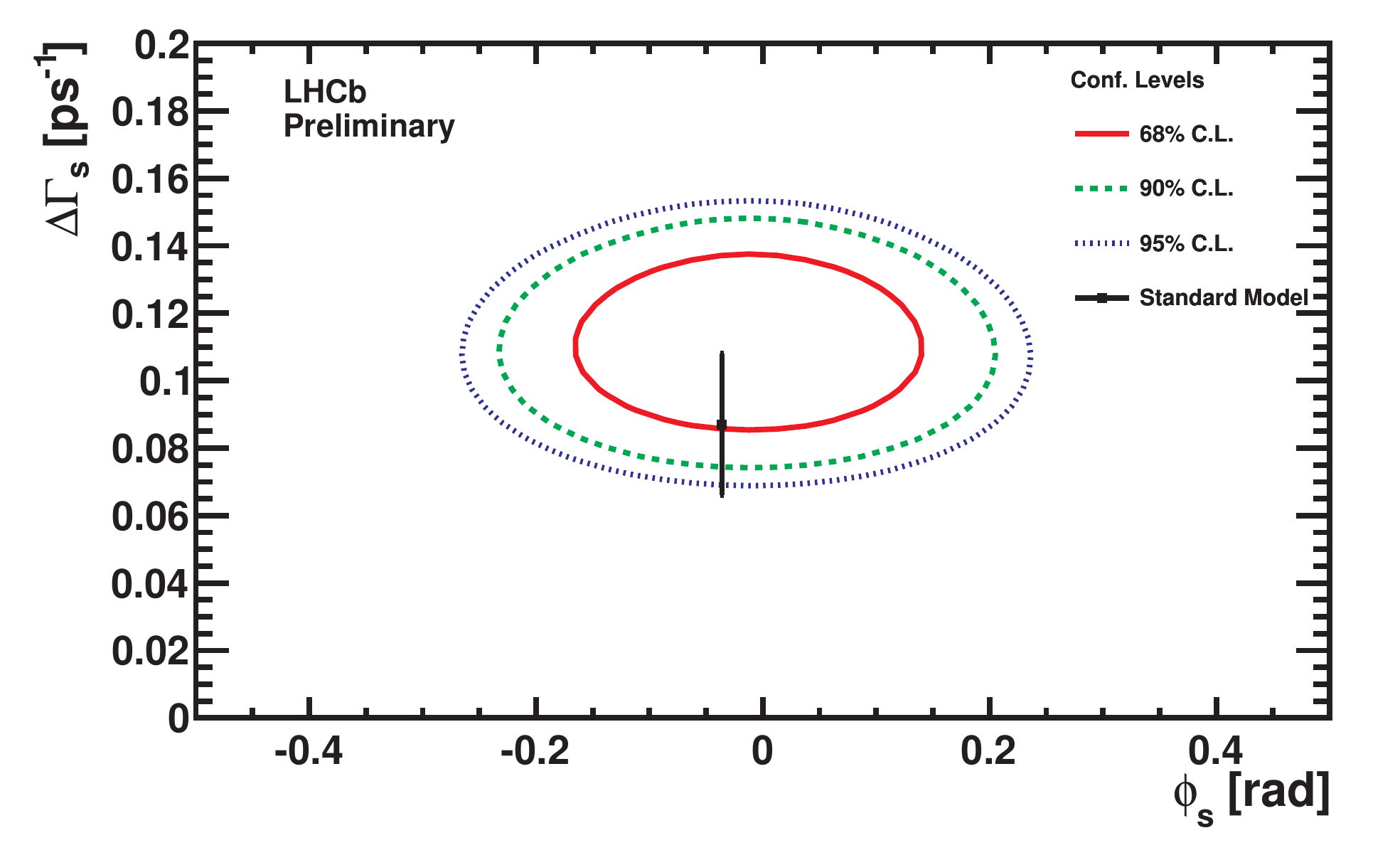}
\caption{Confidence regions in the $\phi_s-\Delta\Gamma_s$ plane for $B_s^0\rightarrow J/\psi\phi$. Only
    statistical uncertainties are included.
    The black square corresponds to the theoretical predicted Standard Model
    value \cite{Charles2011}.\label{fig:LH_scan} }
\end{figure}

The $CP$ violating phase $\phi_s$ is extracting from the  $B_s^0\rightarrow J/\psi\phi$ data with an unbinned maximum likelihood fit to the candidate
invariant mass $m$, the decay time $t$,  the initial flavour of the $B_s^0$ and the 4-body decay angles in the transversity frame $\Omega =\{\cos\theta,\varphi, \cos \psi \}$ defined in Ref.~\cite{Dighe1996}.  The PDFs for signal and background are given in~\cite{PhisLHCbPRL2011}.  Besides $\phi_s$, a set of physics observables are measured. For example, the difference between the heavy and light $B_s^0$ eigenstates, $\Delta\Gamma_s$, the decay width $\Gamma_s$, the polarisation amplitudes  $A_{0}, A_{\perp},A_{\parallel}$ and $A_S$ of the P- and S-wave components of the   of the $K^+K^-$ spectrum.  In the fit, the four different amplitudes, $A_i$, are parameterised by $|A_i(0)|$, the absolute value of the amplitude at $t=0$.   The following normalisation is chosen:  $|A_{0}|^2 + |A_{\perp}|^2 + |A_{\parallel}|^2 = 1$, and the S-wave contribution, $F_S$ is defined as $F_S =  |A_{S}|^2 / (|A_{0}|^2 + |A_{\perp}|^2 + |A_{\parallel}|^2) $. 
Also, the convention $\delta_0=0$ is used. 
This choice of the normalisation is different from the previous analysis. It has been chosen, such that the P-wave amplitudes are independent of the $K^+K^-$ invariant mass range. The $B_s^0$ oscillation frequency was previously  measured at LHCb~\cite{LHCbDeltams} with a very high precision $\Delta m_s = 17.63 \pm 0.11$ ps$^{-1}$. 
This value is used in the fit, where it is allowed to vary within its uncertainties. 
The values obtained for all parameters and the correlation matrix  are given in Table 1 and 2 respectively. Except $\delta_{\parallel} $, all parameters are well behaved and have a parabolic likelihood profile.  In the case of $\delta_{\parallel} $ its central value is close to $\pi$, therefore, appears symmetrically just below $\pi$. The 69\% Confidence Level (C.L) encompasses both minima, and the symmetric 68\% C.L interval $\delta_{\parallel} \in [2.81, 3.47]$ rad is quoted (statistical only). 
The results for $\phi_s$ and $\Delta\Gamma_s$ are in good agreement with the Standard Model prediction quoted in Ref.~\cite{Charles2011}. Figure~\ref{fig:projections}
shows the projection of the PDF on the decay time and the three angles in the transversity basis for candidates in an invariant mass within $\pm$20 MeV/$c^2$ around the 
nominal $B_s^0$ mass. 
\begin{table}[t]
\begin{center}
\begin{tabular}{|c||c|c|c|c|c|} \hline
      & $\Gamma_{\mathrm s}$ & $\Delta\Gamma_{\mathrm s}$ & $|A_{\perp}|^{2}$ & $|A_{0}|^{2}$   & $\phi_s$  \\ \hline\hline
                   $\Gamma_{\mathrm s}$ 		&  1.00 		& $-$0.38 		&  0.39 		& 0.20	& $-$0.01 \\
                   $\Delta\Gamma_{\mathrm s}$ 	&       		&  1.00 		& $-$0.67 		&  0.63 	& $-$0.01\\
                   $|A_{\perp}(0)|^{2}$ 			&       		&       		&  1.00 		& $-$0.53  	& $-$0.01\\
                   $|A_{0}(0)|^{2}$ 			&       		&       		&       		&  1.00 	& $-$0.02\\
                    $\phi_s$					&       		&       		&       		&   		& 1.00\\
\hline \end{tabular}
\end{center}
\caption{Correlation matrix for the statistical uncertainties on $\Gamma_s$, $\Delta \Gamma_s$,
  $|A_{\perp}(0)|^2$, $|A_{0}(0)|^2$ and $\phi_s$. \label{tab:corr}}
\end{table}
Figure~\ref{fig:LH_scan} shows the 68.3\%, 90\% and 95\% profile likelihood confidence level contours in the ($\phi_s-\Delta\Gamma_s$) plane. 
The systematic uncertainties listed in Table~\ref{tab:final+unc.} are those which are not directly treated in the likelihood fit. A breakdown is given in Table~\ref{tab:finalsystematicssummary}.  The uncertainty of $\phi_s$ is dominated by the current imperfect knowledge of the angular acceptances and neglecting the possible 
contributions from direct $CP$ violation. The latter was evaluated based on simulation studies which assumes the $CP$ violation parameter $|\lambda|^2 = 0.95$ or 
$|\lambda|^2 = 1.05$ and the no direct $CP$ violation hypothesis ($|\lambda|^2 = 1$). 
The size of $|\lambda|^2 $ used in this study has been motivated by the fit where $|\lambda|$ is left a free parameter. The uncertainties treated directly in the likelihood fit are those from the tagging calibration parameters, the value of $\Delta m_s$ and the decay time resolution model. 
Their total contributions to the statistical uncertainty on $\phi_s$ is below 5 \%. 

The $CP$ violating phase $\phi_s$ was also measured using $B_s^0\rightarrow J\psi\pi\pi$ decays with an unbinned maximum likelihood fit to the mass, the decay time and the initial flavour of the $B_s^0$. The result is $\phi_s = -0.02\pm 0.17\pm0.02$ rad. This measurement does not require an angular analysis, therefore, systematic arising from the knowledge of the angular acceptances are not present. On the other hand,  the uncertainties due to flavour tagging and resolution are included in 
a similar way to what is done the  $B_s^0\rightarrow J\psi\phi$ analysis. The detailed list of individual systematic uncertainties can be found in Ref.~\cite{PhisLHCbPiPi2012}. 
Both measurements of $\phi_s$ are compatible with each other within uncertainties. They were combined in a simultaneous fit resulting in $\phi_s = -0.002\pm 0.083 \pm 0.027$ rad. 
This analysis results in a twofold ambiguity ($\phi_s \leftrightarrow \pi-\phi_s$ ; $\Delta\Gamma_s\leftrightarrow   - \Delta\Gamma_s$). The ambiguity was resolved in Ref.~\cite{LHCbSignDeltaGammas} by studying the behaviour of the relative phase between the P- and S-wave components of the $K^+K^-$ system. 
The solution with $\Delta\Gamma_s$ is favoured and is the only one quoted in this document.

{\tiny
\begin{table}[t]
  \begin{center}\footnotesize
    \begin{tabular}{|l|c|c|c|c|c|c|c|c|c|}\hline
      Source            & $\Gamma_s$ & $\Delta \Gamma_s$   &
      $A_{\perp}^2$ & $A_{0}^2$ & $F_S$ & $\delta_{\parallel}$ &
      $\delta_{\perp}$
      & $\delta_s$           & $\phi_s$  \\
       & [ps$^{-1}$]&[ps$^{-1}$] & & & & [rad] & [rad] & [rad] & [rad]  \\ \hline
      Description of background      & 0.0010 & 0.004 &- & 0.002 & 0.005 & 0.04 &
      0.04 &
      0.06 & 0.011\\
      Angular acceptances &  0.0018 & 0.002 & 0.012 & 0.024 & 0.005 & 0.12 &
      0.06 & 0.05 & 0.012\\
      $t$ acceptance model            &   0.0062 & 0.002 & 0.001 & 0.001 & - & - &
      - & - & -    \\
      $z$ and momentum scale                &  0.0009 & - & - &- &- &- &- &- &-
      \\
      Production asymmetry ($\pm$ 10\%)        &  0.0002 & 0.002 & - & - & - & - & - & - & 0.008
\\
      CPV mixing \& decay ($\pm$ 5\%) &  0.0003 & 0.002 & - & - & - & - & - & - & 0.020
      \\
      Fit bias & - & 0.001 & 0.003 & - & 0.001 & 0.02 & 0.02 & 0.01 & 0.005\\
       \hline
      Quadratic sum    &  0.0066 & 0.006 & 0.013 & 0.024 & 0.007 & 0.13 & 0.07
      & 0.08 & 0.027\\ \hline
       \end{tabular}
  \end{center}
  \caption{Breakdown and summary of systematic uncertainties for each physics parameter extracted from the unbinned log-likelihood fit.}
  \label{tab:finalsystematicssummary}

\end{table}
}

%%-----------------------------------------------------------------------------
\section{Conclusion}
We have performed a time-dependent angular analysis of 
approximately $21200$ flavour tagged $B_s^0\rightarrow J/\psi\phi$ candidates
obtained from 1fb $^{-1}$ of $pp$ collisions collected
during the 2011 LHCb run at $\sqrt{s}$=7 TeV. We find:
  \[
  \setlength{\arraycolsep}{1mm}
  \begin{array}{ccllllllll}
    \phi_s & \;=\;  & -0.001  & \pm & 0.101  & \mbox{(stat)}&  \pm& 0.027& \mbox{(syst)} & \mbox{(rad)} ,\\
    \Gamma_s &\;=\; & 0.6580  &\pm & 0.0054 & \mbox{(stat)} &\pm & 0.0066 & \mbox{(syst)} & \mbox{ps}^{-1},\rule{0pt}{5mm} \\
    \Delta\Gamma_s    &\;=\; & 0.116   &\pm & 0.018    & \mbox{(stat)} &\pm & 0.006 & \mbox{(syst)} &\mbox{ps}^{-1}.\rule{0pt}{5mm} \\
  \end{array}
  \]

This is the world's most precise measurement of $\phi_s$ and the first direct observation for a 
non-zero value for $\Delta\Gamma_s$. These results are in good agreement with Standard Model
predictions~\cite{Charles2011}. For a combination of this result with an independent analysis of $B_s^0
\rightarrow J/\psi \pi \pi$ decays, we find $\phi_s = -0.002 \pm 0.083 \pm 0.027$ rad.

\end{document}